\documentclass[%
 reprint,
superscriptaddress,
 amsmath,amssymb,
 aps,
 pre
]{revtex4-2}

\usepackage{color}
\usepackage{graphicx}
\usepackage{dcolumn}
\usepackage{bm}

\begin{document}

\preprint{APS/123-QED}

\title{Energetics-based model for a diffusiophoretic motion of a deformable droplet}

\author{Hiroyuki Kitahata}
\email{kitahata@chiba-u.jp}
\affiliation{Department of Physics, Graduate School of Science, Chiba University, Chiba 263-8522, Japan}

\author{Yuki Koyano}
\email{koyano@garnet.kobe-u.ac.jp}
\affiliation{Department of Human Environmental Science, Graduate School of Human Development and Environment, Kobe University, Kobe 657-0011, Japan}

\author{Yasuaki Kobayashi}
\email{ykobayashi@josai.ac.jp}
\affiliation{Research Institute for Electronic Science, Hokkaido University, Sapporo 001-0811, Japan}
\affiliation{Department of Mathematics, Faculty of Science, Josai University, Sakado 350-0295, Japan}

\author{Masaharu Nagayama}
\email{Corresponding author, nagayama@es.hokudai.ac.jp}
\affiliation{Research Institute for Electronic Science, Hokkaido University, Sapporo 001-0811, Japan}

\date{\today}

\begin{abstract}
We construct a mathematical model for a diffusiophoretic motion of a deformable droplet, which is floating on a liquid surface and is driven by the surface tension gradient originating from the surface concentration field of the chemicals that are emitted from the droplet. We define the free energy of the system by including the surface and line energies. From the calculation of the functional of the free energy, we obtain a mathematical model for the diffusiophoretic motion with deformation. By only considering the deformation of the second mode, we explicitly derive the time-evolution equations for the translational motion and the elliptic deformation. There are three stable states: an immobile circular droplet, an immobile elliptically deformed droplet, and a mobile droplet with the elliptic deformation in which the minor axis meets the motion direction, and we discuss the transition between these three stable states.
\end{abstract}

\maketitle
  
\section{Introduction \label{sec:introduction}}

Under nonequilibrium conditions, particles and droplets can spontaneously move through the dissipation of chemical energy, which is often called self-propulsion. They have been intensively studied as a model for the motion of living organisms. They have also attracted interest as elements of active matter, in which spatio-temporal order at greater scales than those of elements emerges~\cite{Viscek_2012, Bechinger_RMP, Maass}. Many experimental systems on self-propulsion in physico-chemical systems have also been reported; crawling droplets on a solid substrate~\cite{dosSantos1995,DEGENNES1999147,Lee_PhysRevE.65.051602,Magome1996,Sumino_2005}, oil droplets in surfactant aqueous solutions or aqueous droplets in oil with surfactant exhibit self-propulsion~\cite{toyota2009JACS,Izri_PRL2014,Thutupalli_2011,yamamoto_SoftMatter, Herminghaus_SoftMatter2014, Suga_Kimura_2018, Michelin, Michelin_AnnuRev}. Other examples are Quincke rollers, self-propelled plastic particles under an electric field~\cite{Bricard2013, Kato_SoftMatter}, and Janus particles, two-faced particles that move due to the temperature gradient or chemical potential gradient~\cite{Paxton2004, Howse_2007, Jiang_Sano_PRL2010, Ke2010}. Floating objects on liquid surfaces can exhibit self-propulsion due to the difference in surface tension, which are often called Marangoni surfers~\cite{Nakata_Langmuir1997, PCCP_camphor_review2015, Nakata2003JPCB, Taylor_JPCB2013, AKELLA20181176}. Droplet motions with adsorption and desorption of surfactants were discussed based on the low-Reynolds number hydrodynamics~\cite{Yabunaka_Ohta_Yoshinaga_2012,Yoshinaga_PRE_2012,Yoshinaga_PRE2014}, and the self-propelled droplet on the solid surface in a surfactant aqueous solution is discussed based on the energetics~\cite{Thiele_PRL,Brochard1989,Thiele_PRL,Thiele_PhysRevE2020}. The solid Marangoni surfers have also been discussed based on hydrodynamics~\cite{Lauga_Davis_2012,Gouiller_PRE2021} or the coupled system between Newtonian equations and reaction-diffusion equations~\cite{HayashimaJPCB2001,Nagayama_PhysicaD,PCCP_camphor_review2015,Springer_Book_Chap2}.

In self-propulsion systems, the coupling between motion and shape has attracted interest. This is partly because the motion of living cells such as \textit{Keratocyte} and \textit{Dictyostelium} is strongly correlated with their shapes~\cite{keren2008mechanism,Bosgraaf_PlosONE2009,ebata2024amoeboid}. 
In physico-chemical systems, the shape-dependent self-propulsion has also been reported. Droplets that exhibit Marangoni surfing are often deformed and their shapes are strongly correlated with the direction of the motion~\cite{Nagai_PRE2005,Nagai_PTPS,Bates_Langmuir_2008,STanaka_PRE_2015,STanaka_JPSJ2017,Pimienta_ACIE2011,BONIFACE2022990,Loeffler,otani2024deformation}. This is because both self-propulsion and deformation have the same origin. For example, an oil droplet in a surfactant aqueous solution exhibits a self-propelled motion due to the Marangoni convection, which also causes ellipsoidal deformation~\cite{banno2016deformable}.
Gel formation at the interface is also known to cause motion and deformation of the droplet~\cite{Sumino_gel_PRE2007} and a similar droplet behavior due to the actomyosin force generation was reported recently~\cite{Maeda_PNAS2022}. From a viewpoint of theoretical approach to the coupling between self-propulsion and deformation, there has also been a wide variety of mathematical modeling. One of the simplest models, so-called Ohta-Ohkuma model, described by coupled ordinary differential equations is proposed and analyzed based on the system symmetry~\cite{Ohta_Ohkuma2009, Tarama_2016, Ohta_JPSJ2017}. This model only suggests the possible coupling manner and does not include the correspondence to the mechanism of the actual systems. On the other hand, some models have been developed based on insights from actual phenomena of the droplet motion and deformation~\cite{Thiele_PRL, Nagai_JCP2016, Nagayama2023_SciRep}. Mathematical models that reproduce the motion and deformation of the cells have also been proposed using the phase-field model~\cite{Ziebert_pone,Ziebert2016,Levine_PRL2010} and the Fourier series expansion~\cite{saito_ishihara}.

We have investigated the motion of Marangoni surfers based on experiments and mathematical modeling comprising the Newtonian equations and reaction-diffusion equations. The Marangoni surfers are driven by surface tension gradient originating from the concentration profile of chemicals that are emitted from the surfers. The surfers tend to escape from the chemicals and thus their motion mechanism is often called self-diffusiophoresis. We have considered such a self-diffusiophoretic particle with a given shape that emits surface-active chemicals and moves due to the surface tension gradient~\cite{Kitahata_Iida_2013,iida2014theoretical,Kitahata_JPSJ_2020,Kitahata_Frontier2022,Springer_Book_Chap2}. For example, theoretical analysis suggests that an elliptic Marangoni surfer should move in its minor axis direction, which is also confirmed by numerical simulations and experiments using a camphor particle. In these studies, the surfer has a rigid shape, and thus we do not need to discuss the change in shape. In contrast, deformable droplets can also move as Marangoni surfers. For example, a pure alcohol droplet like pentanol and hexanol on an alcohol aqueous solution is known as the examples of such deformable moving droplets~\cite{Nagai_PRE2005,Nagai_PTPS,Oshima2014}. These experimental results clearly demonstrate the coupling between motion and deformation. Some of the authors proposed a mathematical model based on the energy minimization but the cost for numerical simulation was quite high~\cite{Nagai_JCP2016}. Therefore, we await a mathematical model that well reproduces experimental results and enables us to discuss the mechanism from a viewpoint of system symmetry. Recently, some of the authors proposed a mathematical model adopting a phase-field description for the droplet shape~\cite{Nagayama2023_SciRep}. The model succeeded in reproducing self-propulsion coupled with deformation but the relation to the simple models like the Ohta-Ohkuma model~\cite{Ohta_Ohkuma2009,Tarama_2016}, is still to be unveiled. Through the discussion of the correspondence between these models, we believe that we can attain the essential mechanism of the diffusiophoretic motion with deformation.

In the present study, thus, we propose a mathematical model for the diffusiophoretic deformable droplet, in which a droplet shape is described by the Fourier series expansion. Our model corresponds to a system with a floating deformable droplet that releases surface-active chemicals and generates their concentration field. The motion and deformation of the droplet are driven by the surface tension gradient originating from the concentration field of the chemicals. The system dynamics are partly derived using a variational principle of the free energy of the system. In the manuscript, we first show our mathematical model together with brief derivation of it in Sect.~\ref{sec:modeling}. Then, we limit the case in which a droplet shows translational motion and the 2-mode deformation. We show the explicit form of the model and show the numerical simulation results in Sect.~\ref{sec:simulation}. Further, we adopt the perturbation method and obtain the reduced model in Sect.~\ref{sec:theory}. Then, we perform a linear stability analysis of the obtained reduced model and compare with numerical simulation results in Sect.~\ref{sec:ODE} Finally, we summarize our work and show the future direction in Sect.~\ref{sec:discussion}. Since the present model is derived naturally from the free energy of the system, the model can be a universal one for the diffusiophoretic motion of a deformable object in a two-dimensional space.

\section{Modeling \label{sec:modeling}}

We consider a two-dimensional system in which a deformable droplet is floating on a liquid surface and is driven by the surface tension gradient originating from the concentration profile of chemicals emitted from the droplet. It should be noted that we do not consider the hydrodynamics or depth of the system, but only consider a two-dimensional system. The deformation of a droplet is described by the deviation from a circular shape. We consider the local polar coordinates $(r, \theta)$ where the origin meets the center of mass $\bm{r}_c$ of the droplet. Then, the position of the droplet periphery is expressed as
\begin{align}
r =& f(\theta; \left\{a_k\right\}, \left\{ b_k \right\} ) \nonumber \\
=&R \left[1 + \sum_{k=2}^\infty \left(a_k \cos k\theta + b_k \sin k\theta \right) \right],
\end{align}
where $R$ is the droplet radius. The coefficients $\left \{ a_k \right \}$ and $\left \{ b_k \right \}$ are on the order of a small parameter $\epsilon$, and indicate the deformation amplitudes for the $k$th mode. Here, we only consider the small deformation from a circle that can be described by the first order of the small parameter $\epsilon$. It is notable that the 0th and 1st modes are omitted since they correspond to the expansion/contraction and translation, respectively. 
The region inside the deformable droplet is given as
\begin{align}
\Omega( \bm{r}_c, \left\{a_k\right\}, \left\{b_k\right\}) = \left\{  \bm{r} \middle| \bm{r} - \bm{r}_c \in \Omega_0(\left\{a_k\right\}, \left\{b_k\right\})  \right\}, \label{Omega}
\end{align}
with
\begin{align}
\Omega_0( \left\{a_k\right\}, \left\{b_k\right\}) = \left\{  \bm{r} = r \bm{e}_r(\theta) \middle| r \leq f(\theta; \left\{a_k\right\}, \left\{ b_k \right\} ) \right\}, \label{Omega0}
\end{align}
where $\bm{e}_r(\theta)$ and $\bm{e}_\theta(\theta)$ are the unit vectors in the polar coordinates, which are explicitly defined as $\bm{e}_r(\theta) = \cos \theta \bm{e}_x + \sin \theta \bm{e}_y$ and $\bm{e}_\theta(\theta) = -\sin \theta\bm{e}_x + \cos \theta \bm{e}_y$ using the unit vectors in the Cartesian coordinates.

Considering that the droplet emits surface-active chemicals, and the emitted chemicals exhibit diffusion at the water surface and evaporate to the air phase or dissolve into a bulk aqueous phase, we consider the time evolution of the concentration $u$ of the chemicals obeys the reaction-diffusion equation as
\begin{align}
\frac{\partial u}{\partial t} = D \nabla^2 u - \alpha u + S(\bm{r} - \bm{r}_c; \left\{a_k\right\}, \left\{b_k\right\}). \label{RD}
\end{align}
Here, $\alpha$ is the evaporation and dissolution rate. The transport of chemicals should be induced by the Marangoni effect~\cite{SCRIVEN1960} and it should be described by a nonlinear term. However, it is shown that the spreading process of the concentration field by the Marangoni convection can be approximated using a linear diffusion term $D\nabla^2 u$ with an effective diffusion coefficient $D$ if the system is close to the bifurcation point from a steady state~\cite{JCP_HK_NY,suematsu2014quantitative,Bickel_PRE}. Thus, we adopt the effective linear diffusion term. $S(\bm{r}; \left\{a_k\right\}, \left\{b_k\right\})$ are the supply rate of the surface-active chemicals from the droplet. The term $S$ is explicitly expressed as
\begin{align}
S(\bm{r}; \left\{ a_k\right\}, \left\{ b_k \right\}) =  \left\{ \begin{array}{ll} S_0 / A, & \bm{r} \in \Omega_0(\left\{a_k\right\}, \left\{b_k\right\}), \\ 0, & \bm{r} \notin \Omega_0(\left\{a_k\right\}, \left\{b_k\right\}). \end{array} \right.
\end{align}
Here, $S_0$ is the supply rate of the surface active chemicals from the droplet, and $A$ is the base area of the droplet, which is explicitly obtained as $ A = \pi R^2 + \mathcal{O}(\epsilon^2)$. 
In our model, we assume that the chemicals are released not only from the droplet periphery but also from the whole droplet region. If we only introduce the supply of chemicals at the periphery, we need to consider the reaction-diffusion equation that is valid only for the outside of the droplet region. In such a case, we need to include a Neumann boundary condition at the moving droplet periphery. By introducing the approximation that the chemicals are supplied in the whole droplet region, we obtain a time evolution equation that is valid for the whole two-dimensional space and can approximate the concentration dynamics around the droplet.

In order to derive the time-evolution equations for the droplet velocity and deformation, we consider the free energy of the system as
\begin{align}
E =& E_s  + E_l
\end{align}
where the first term $E_s$ on the right-hand side denotes the surface energy depending on the surface tension $\gamma$ as a function of the local chemical concentration $u$ as
\begin{align}
E_s = \iint_{\mathbb{R}^2 \backslash \Omega(\bm{r}_c, \left\{a_k\right\}, \left\{b_k\right\})} \gamma \left(u( \bm{r}) \right)  dA,
\end{align}
and the second term $E_l$ denotes the line energy proportional to the peripheral length of the droplet,
\begin{align}
    E_l = \tau \oint_{\partial \Omega(\bm{r}_c, \left\{a_k\right\}, \left\{b_k\right\})} d\ell.
\end{align}
Here, $dA$ and $d\ell$ denote the area and line elements, respectively, and $\partial \Omega$ denotes the periphery of the region $\Omega$.
It is known that the line tension is negligibly small compared with the surface tension, but if the droplet is flat enough except for its periphery, it has extra energy which should be proportional to the peripheral length~\cite{de2003capillarity,otani2024deformation}. We thus assume that $\tau$ is a positive constant.

The time evolution of the droplet velocity and deformation is given as
\begin{align}
\eta_t \frac{d \bm{r}_c}{dt} = - \frac{\partial E}{\partial \bm{r}_c}, \label{drcdt}
\end{align}
\begin{align}
\eta_k \frac{d a_k}{dt} = - \frac{\partial E}{\partial a_k},
\label{da2dt}
\end{align}
\begin{align}
\eta_k \frac{d b_k}{dt} =  -\frac{\partial E}{\partial b_k},
\label{db2dt}
\end{align}
for $k = 2, 3, 4, \dots$, where $\eta_t$ and $\eta_k$ are the positive coefficients to determine the time scale of the translational motion and deformation, respectively. The right sides of these equations can be calculated as
\begin{align}
\frac{\partial E_s}{\partial \bm{r}_c}
&= - \iint_{\Omega(\bm{r}_c, \left\{ a_k \right\}, \left\{ b_k\right\})} \nabla \gamma(u(\bm{r})) dA\nonumber \\
&= -\oint_{\partial \Omega(\bm{r}_c, \left\{ a_k \right\}, \left\{ b_k\right\})} \gamma(u(\bm{r})) \bm{e}_n(\bm{r}) d\ell,
\label{dedrc}
\end{align}
\begin{align}
\frac{\partial E_s}{\partial a_k} =& - \iint_{\Omega \left(\bm{r}_c, \left\{ a_k \right\} , \left\{ b_k \right\} \right)}\bm{w}^{(a)}_k \cdot \nabla \gamma(u(\bm{r})) dA \nonumber \\
=& - \oint_{\partial \Omega \left(\bm{r}_c, \left\{ a_k \right\} , \left\{ b_k \right\} \right)} \gamma(u(\bm{r})) \bm{w}^{(a)}_k \cdot \bm{e}_n(\bm{r}) d \ell, \label{dedak}
\end{align}
\begin{align}
\frac{\partial E_s}{\partial b_k} =& - \iint_{\Omega \left(\bm{r}_c, \left\{ a_k \right\} , \left\{ b_k \right\} \right)} \bm{w}^{(b)}_k \cdot \nabla \gamma(u(\bm{r})) dA \nonumber \\
=& - \oint_{\partial \Omega \left(\bm{r}_c, \left\{ a_k \right\} , \left\{ b_k \right\} \right)} \gamma(u(\bm{r})) \bm{w}^{(b)}_k \cdot \bm{e}_n(\bm{r}) d \ell.
\label{dedbk}
\end{align}
Here, we introduce incompressible vector fields $\bm{w}^{(a)}_k$ and $\bm{w}^{(b)}_k$, whose explicit expressions are given as
\begin{align}
\bm{w}^{(a)}_k =& \nabla \left[ \frac{R^2}{k} \left(\frac{r}{R} \right)^k \cos k\theta \right] \nonumber \\
=& R \left( \frac{r}{R} \right)^{k-1} \left( \cos k\theta \bm{e}_r(\theta) - \sin k\theta\bm{e}_\theta (\theta) \right), \label{wka}
\end{align}
\begin{align}
\bm{w}^{(b)}_k =& \nabla \left[ \frac{R^2}{k} \left(\frac{r}{R} \right)^k \sin k\theta \right] \nonumber \\
=& R \left( \frac{r}{R} \right)^{k-1} \left( \sin k\theta \bm{e}_r(\theta) + \cos k\theta\bm{e}_\theta (\theta) \right). \label{wkb}
\end{align}
It should be noted that $\bm{w}_k^{(a)}$ and $\bm{w}_k^{(b)}$ correspond to the incompressible flow field that obeys the Stokes equation. The detailed calculation is shown in Appendix~\ref{App_C}.

As for the extra surface energy for the periphery, we obtain
\begin{align}
E_l &= 2 \pi \tau R \left[ 1 + \frac{1}{4} \sum_{k=2}^{\infty} k^2 \left ( {a_k}^2 + {b_k}^2 \right ) \right] + \mathcal{O}(\epsilon^3).
\end{align}
Therefore,
\begin{align}
\frac{\partial E_l}{\partial \bm{r}_c} = 0,
\end{align}
\begin{align}
\frac{\partial E_l}{\partial a_k} =  \pi k^2 \tau R a_k +  \mathcal{O}(\epsilon^2) = \kappa_k a_k +  \mathcal{O}(\epsilon^2),
\end{align}
\begin{align}
\frac{\partial E_l}{\partial b_k} = \pi k^2 \tau R b_k  + \mathcal{O}(\epsilon^2) =  \kappa_k b_k +  \mathcal{O}(\epsilon^2),
\end{align}
where $\kappa_k$ is defined as
\begin{align}
\kappa_k = \pi k^2 \tau R > 0.
\end{align}

Thus, we finally obtain
\begin{align}
\eta_t \frac{d \bm{r}_c}{dt} 
&= \iint_{\Omega(\bm{r}_c, \left\{ a_k \right\}, \left\{ b_k\right\})} \nabla \gamma(u(\bm{r})) dA \nonumber \\
&= \oint_{\partial \Omega(\bm{r}_c, \left\{ a_k \right\}, \left\{ b_k\right\})} \gamma(u(\bm{r})) \bm{e}_n(\bm{r}) d\ell ,
\end{align}
\begin{align}
\eta_k \frac{d a_k}{dt} 
=& \iint_{\Omega \left(\bm{r}_c, \left\{ a_k \right\} , \left\{ b_k \right\} \right)} \bm{w}^{(a)}_k \cdot \nabla \gamma
(u(\bm{r})) dA - \kappa_k a_k \nonumber \\
=& \oint_{\partial \Omega \left(\bm{r}_c, \left\{ a_k \right\} , \left\{ b_k \right\} \right)} \gamma(u(\bm{r})) \bm{w}^{(a)}_k \cdot \bm{e}_n(\bm{r}) d \ell - \kappa_k a_k,
\end{align}
\begin{align}
\eta_k \frac{d b_k}{dt} 
=& \iint_{\Omega \left(\bm{r}_c, \left\{ a_k \right\} , \left\{ b_k \right\} \right)} \bm{w}^{(b)}_k \cdot \nabla \gamma(u(\bm{r})) dA - \kappa_k b_k
\nonumber \\
=& \oint_{\partial \Omega \left(\bm{r}_c, \left\{ a_k \right\} , \left\{ b_k \right\} \right)} \gamma(u(\bm{r})) \bm{w}^{(b)}_k \cdot \bm{e}_n(\bm{r}) d \ell - \kappa_k b_k  .
\end{align}
The first line in each equation is suitable for the numerical simulation while the second line is suitable for the theoretical analysis.

The relationship between the surface tension and concentration of the chemical is assumed to be
\begin{align}
\gamma = \gamma_0 - \Gamma u,
\end{align}
where $\gamma_0$ is the surface tension of pure water and $\Gamma$ is a positive constant.

Hereafter, we consider the dimensionless version of the model, where the scales for length, time, concentration, and force are set as $\sqrt{D/\alpha}$, $1/\alpha$, $S_0/\alpha$, $\Gamma S_0 / \sqrt{\alpha D}$, respectively. Then, the dimensionless version of our model is as follows
\begin{align}
\eta_t \frac{d \bm{r}_c}{dt} 
&= -\iint_{\Omega(\bm{r}_c, \left\{ a_k \right\}, \left\{ b_k\right\})} \nabla u(\bm{r}) dA \nonumber \\
&= -\oint_{\partial \Omega(\bm{r}_c, \left\{ a_k \right\}, \left\{ b_k\right\})} u(\bm{r}) \bm{e}_n(\bm{r}) d\ell , \label{drcdt_dl}
\end{align}
\begin{align}
\eta_k \frac{d a_k}{dt} 
=& - \iint_{\Omega \left(\bm{r}_c, \left\{ a_k \right\} , \left\{ b_k \right\} \right)} \bm{w}^{(a)}_k \cdot \nabla 
u(\bm{r}) dA - \kappa_k a_k \nonumber \\
=& - \oint_{\partial \Omega \left(\bm{r}_c, \left\{ a_k \right\} , \left\{ b_k \right\} \right)} u(\bm{r}) \bm{w}^{(a)}_k \cdot \bm{e}_n(\bm{r}) d \ell - \kappa_k a_k , \label{dakdt_dl}
\end{align}
\begin{align}
\eta_k \frac{d b_k}{dt} 
=& - \iint_{\Omega \left(\bm{r}_c, \left\{ a_k \right\} , \left\{ b_k \right\} \right)} \bm{w}^{(b)}_k \cdot \nabla u(\bm{r}) dA - \kappa_k b_k 
\nonumber \\
=& - \oint_{\partial \Omega \left(\bm{r}_c, \left\{ a_k \right\} , \left\{ b_k \right\} \right)} u(\bm{r}) \bm{w}^{(b)}_k \cdot \bm{e}_n(\bm{r}) d \ell - \kappa_k b_k , \label{dbkdt_dl}
\end{align}
\begin{align}
\frac{\partial u}{\partial t} = \nabla^2 u - u + S(\bm{r}-\bm{r}_c ; \left\{a_k\right\}, \left\{b_k\right\}), \label{eq_u_dl}
\end{align}
\begin{align}
S(\bm{r}; \left\{ a_k\right\}, \left\{ b_k \right\}) =  \left\{ \begin{array}{ll} 1 / A, & \bm{r} \in \Omega_0(\left\{a_k\right\}, \left\{b_k\right\}), \\ 0, & \bm{r}  \notin \Omega_0(\left\{a_k\right\}, \left\{b_k\right\}), \end{array} \right. \label{S_def}
\end{align}
with Eqs.~\eqref{Omega} and \eqref{Omega0}.

\section{Numerical simulation in the case with 2-mode deformation \label{sec:simulation}}

Based on the partial differential equation (PDE) model derived above, we perform a numerical simulation. For the correspondence to the theoretical analyses described in the following section, we only consider the 2-mode deformation. In order to minimize the effect of discretization, we define a smooth function representing the area inside the droplet as
\begin{align}
    \Theta_\sigma(\bm{r}; a_2,b_2) = 
& \frac{1}{2} \left[ 1 + \tanh \left( - \frac{r - f(\theta; a_2, b_2) }{\sigma}\right) \right], \label{eq31}
\end{align}
where
\begin{align}
   f(\theta; a_2, b_2) = R\left(1 + a_2 \cos2\theta + b_2 \sin2\theta\right). 
\end{align}
Here, $\sigma$ is a positive constant for the smooth connection of variables at the boundary. As for the calculation of the time evolution of $\bm{r}_c$, $a_2$, and $b_2$, we adopt $\Theta_\sigma(\bm{r}_c; a_2, b_2)$ as
\begin{align}
\eta_t \frac{d \bm{r}_c}{dt} 
= -\iint_{\mathbb{R}^2} \nabla u(\bm{r}) \Theta_\sigma(\bm{r}-\bm{r}_c; a_2,b_2) dA,
\end{align}
\begin{align}
\eta_2 \frac{d a_2}{dt} 
= - \iint_{\mathbb{R}^2} \bm{w}^{(a)}_2 \cdot \nabla 
u(\bm{r}) \Theta_\sigma(\bm{r}-\bm{r}_c; a_2,b_2) dA - \kappa_2 a_2,
\end{align}
\begin{align}
\eta_2 \frac{d b_2}{dt} 
=& - \iint_{\mathbb{R}^2} \bm{w}^{(b)}_2 \cdot \nabla u(\bm{r}) \Theta_\sigma(\bm{r}-\bm{r}_c;a_2,b_2) dA - \kappa_2 b_2 ,
\end{align}
\begin{align}
\frac{\partial u}{\partial t} = \nabla^2 u - u + S_\sigma(\bm{r}-\bm{r}_c ; a_2, b_2), \end{align}
where
\begin{align}
    S_\sigma(\bm{r}; a_2, b_2) = \frac{1}{A} \Theta_\sigma(\bm{r}; a_2, b_2).
\end{align}
 It should be noted that $S_\sigma(\bm{r};a_2,b_2)$ converges to $S(\bm{r};\left\{a_k\right\},\left\{b_k\right\})$ in Eq.~\eqref{S_def} with $a_3 = a_4 = \dots = 0$ and $b_3 = b_4 = \dots = 0$ when $\sigma \to +0$. $\bm{w}_2^{(a)}$ and $\bm{w}_2^{(b)}$ are explicitly given as
 \begin{align}
\bm{w}_2^{(a)} = x \bm{e}_x - y \bm{e}_y, 
 \end{align}
\begin{align}
\bm{w}_2^{(b)} = y \bm{e}_x + x \bm{e}_y. \label{eq39}
 \end{align}

For the numerical simulation, we adopt an explicit method for the time development with spatial mesh $\Delta x = 0.025$ and time step $\Delta t = 0.0001$. The system size is set to be $7 \times 7$ with the periodic boundary condition. The smoothing parameter is set to be $\sigma = 0.03$. It is confirmed that the results do not change significantly if we slightly change the value of $\sigma$. We fix $R=1$ and $\eta_2 = 0.1$, and vary $\eta_t$ and $\kappa_2$ as the parameters. The initial conditions for $u$ and $\bm{r}_c$ are fixed as $u = 0$ and $\bm{r}_c = \bm{0}$, and those for $d\bm{r}_c/dt$, $a_2$, and $b_2$ are varied.
\begin{figure}
    \centering
    \includegraphics{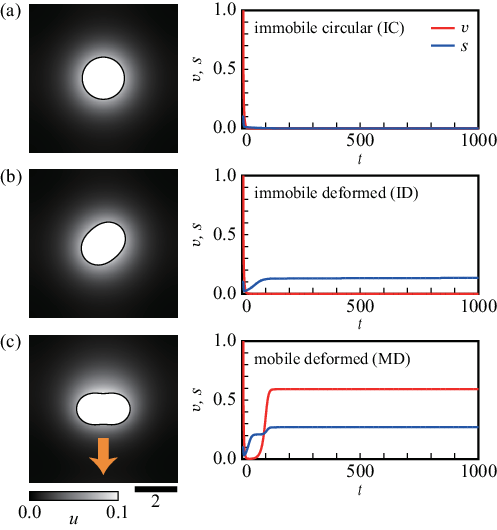}
    \caption{Numerical simulation results obtained from the PDE model in Eqs.~\eqref{eq31} to \eqref{eq39}. Snapshots at $t = 1000$ (left panels) and time series (right panels) of the speed $v = \left|d\bm{r}_c/dt\right|$ (red) and the deformation magnitude $s = \sqrt{{a_2}^2+{b_2}^2}$ (blue) are shown for the three typical cases. (a) Immobile circular (IC) droplet at $\kappa_2 = 0.12$ and $\eta_t = 0.085$. (b) Immobile deformed (ID) droplet at $\kappa_2 = 0.116$ and $\eta_t = 0.085$. (c) Mobile deformed (MD) droplet at $\kappa_2 = 0.112$ and $\eta_t = 0.085$. The droplet moves downward as the orange arrow indicates. Grayscale tone displays the concentration field $u$. The initial condition is set as $d\bm{r}_c/dt = \bm{e}_x + 0.1 \bm{e}_y$, $a_2 = 0.1$, $b_2 = 0.01$.}
    \label{fig1}
\end{figure}

In Fig.~\ref{fig1}, we show a typical example of the droplet dynamics. For great $\eta_t$ and $\kappa_2$, we observe an immobile circular (IC) droplet, which keeps a circular shape and does not move as shown in Fig.~\ref{fig1}(a). For smaller $\kappa_2$ and greater $\eta_t$, we observe an immobile deformed (ID) droplet, which deforms into an elliptic or peanut shape but does not move, as shown in Fig.~\ref{fig1}(b). For smaller $\eta_t$, we observe a mobile deformed (MD) droplet, which moves in a certain direction with an elliptic deformation whose direction of the minor axis meets the direction of motion as shown in Fig.~\ref{fig1}(c). 

\begin{figure*}
    \centering
    \includegraphics{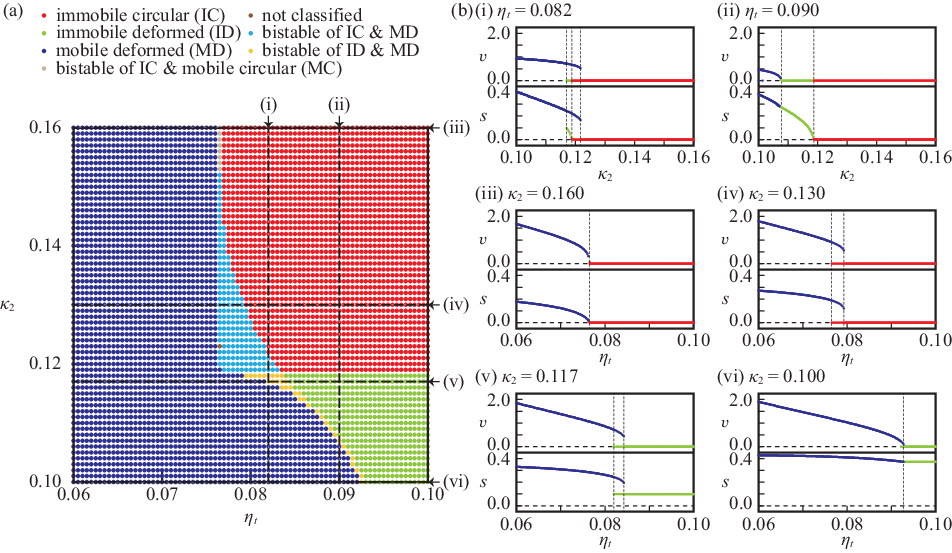}
    \caption{Numerical simulation results obtained from the PDE model in Eqs.~\eqref{eq31} to \eqref{eq39}. (a) Two-dimensional phase diagram on the $\eta_t$-$\kappa_2$ plane. The other parameters are set to be $\eta_2 = 0.1$ and $R=1$. The red, green, and dark blue points correspond to an immobile circular (IC) droplet, an immobile deformed (ID) droplet, and a mobile deformed (MD) droplet. The cyan points show the bistability between IC and MD droplets, the yellow points show that between ID and MD droplets, and the gray points show that between IC and mobile circular (MC) droplet. The brown points represent the case that we can classify into none of the above (three-or-more final steady states due to the slow convergence of the system). (b) One-dimensional bifurcation diagrams with constant $\eta_t$ or $\kappa_2$, which are indicated as arrows at the top and right sides of the plot in panel (a). Speed $v$ and deformation $s$ are plotted. (i,ii) $\kappa_2$ is varied at (i) $\eta_t = 0.082$ and (ii) $\eta_t = 0.090$. (iii-vi) $\eta_t$ is varied at (iii) $\kappa_2 = 0.160$, (iv) $\kappa_2 = 0.130$, (v) $\kappa_2 = 0.117$, and (vi) $\kappa_2 = 0.100$.  The red, green, and dark blue lines represent immobile circular (IC), immobile deformed (ID), and mobile deformed (MD) droplets. For the MD droplet, it moves in its minor-axis direction. The vertical dotted lines are drawn to indicate the transition points and correspondence between $v$ and $s$.}
    \label{fig2}
\end{figure*}

In Fig.~\ref{fig2}(a), we show a two-dimensional phase diagram of stable states on the $\eta_t$-$\kappa_2$ plane. The other parameters are fixed at $\eta_2 = 0.1$ and $R=1$. We start with the following four initial conditions (i) $d\bm{r}_c/dt = \bm{e}_x + 0.1 \bm{e}_y$, $a_2 = 0.1$, $b_2 = 0.01$, (ii) $d\bm{r}_c/dt = \bm{e}_x + 0.01 \bm{e}_y$, $a_2 = -0.25$, $b_2 = 0.002$, (iii) $d\bm{r}_c/dt = 0.01 \bm{e}_x + 0.001 \bm{e}_y$, $a_2 = 0.1$, $b_2 = 0.01$, and (iv) $d\bm{r}_c/dt = 0.01\bm{e}_x + 0.001 \bm{e}_y$, $a_2 = 0.01$, $b_2 = 0.001$, where the initial conditions (i) and (ii) have a large velocity and deformation with different direction relations, (iii) has a small velocity and a large deformation, and (iv) has a small velocity and deformation. From each initial condition, the numerical simulation is run until it reaches a steady state, where we consider the system reaches a steady state if maximum time difference of $v_x$, $v_y$, $a_2$, and $b_2$ are less than $10^{-6}$ per 0.1 time unit. We classify the final steady-state using $v = \sqrt{{v_x}^2 + {v_y}^2}$ and $s = \sqrt{{a_2}^2 + {b_2}^2}$ as follows: an IC droplet in case that $v \leq 0.01$ and $s \leq 0.01$, an ID droplet in case that $v \leq 0.01$ and $s > 0.01$, an MD droplet in case that $v > 0.01$ and $s > 0.01$, and a mobile circular (MC) droplet in case that $v>0.01$ and $s\leq 0.01$. If the final steady states are the same from all four initial conditions, then we consider the system to have a single final state. If there are two final steady states depending on the choice of the initial conditions, then the system is considered to be in a bistable state. If there are three-or-more final steady states, then the system is ``not classified''.

In the phase diagram, the IC droplet is stable for $\eta_t  \gtrsim  0.077$
 and $\kappa_2\gtrsim 0.119$. With a decrease in $\kappa_2$, the IC droplet becomes unstable and an ID droplet becomes stable. In contrast, with a decrease in $\eta_t$, an IC droplet becomes unstable and an MD droplet is realized. Close to the transition points between IC and MD droplets and those between ID and MD droplets, we observe the bistable states of them, respectively.
 
The gray points around $\eta_t \simeq 0.077$ and $\kappa_2 \gtrsim 0.153$ correspond to the bistable state between the IC droplet and the mobile circular (MC) droplet, where the droplet has a finite speed with a small deformation. This steady state is classified into the MC droplet since the amplitude of deformation is quite small. Thus, we consider that the states corresponding to the gray points are the intermediate region between IC and MD droplets.

In Fig.~\ref{fig2}(b), the one-dimensional bifurcation diagrams are shown. Here, we adopt symmetric initial conditions (i') $d\bm{r}_c/dt = \bm{e}_x$, $a_2 = -0.1$, $b_2 = 0.0$, and (ii') $d\bm{r}_c/dt =0.001 \bm{e}_x$, $a_2 = -0.1$, $b_2 = 0$ to clearly show the bifurcation structures of the solutions. In Fig.~\ref{fig2}(b)-(i,ii), the bifurcation diagrams with respect to $\kappa_2$ at $\eta_t=0.082$(i) and $0.090$(ii) are shown. For $\eta_t = 0.082$, the bistability of the MD and ID droplets and that of the MD and IC droplets are shown. The transition between the IC and ID droplets is continuous. For $\eta_t = 0.090$, the IC droplet continuously changes to the ID droplet and then to the MD droplet with a decrease in $\kappa_2$. In Fig.~\ref{fig2}(b)-(iii-vi), the bifurcation diagrams with respect to $\eta_t$ for $\kappa_2 = 0.160$(iii), $0.130$(iv), $0,117$(v) and $0.100$(vi) are shown. The IC droplet continuously changes to the MD droplet with a decrease in $\eta_t$ for $\kappa_2 = 0.160$, while the transition between the IC and MD droplets is discontinuous and has a hysteresis for $\kappa_2 = 0.130$. Also for the transition between the ID and MD droplets, discontinuous and continuous transitions are seen for $\kappa_2 = 0.117$ and $0.100$, respectively.

For even smaller $\eta_t$ or $\kappa_2$, the calculation can diverge due to $s$. This is because our modeling is based on the assumption that the deformation is sufficiently small as the first-order perturbation from a circle. Actually, $s$ is well-defined in the range of $\left|s\right| < 1$. 

\section{Reduction to ODE model including only 2-mode deformation \label{sec:theory}}

Here, we also consider a droplet motion only with the 2-mode deformation. The driving force for translational motion $\bm{F}$ and those for deformation $G^{(2a)}$ and $G^{(2b)}$ that originate from $E_s$ are defined as 
\begin{align}
    \bm{F} =  -\oint_{\partial \Omega \left(\bm{r}_c, a_2, b_2 \right)} u(\bm{r}) \bm{e}_n(\bm{r}) d \ell, \label{Fdefinition}
\end{align}
\begin{align}
    G^{(2a)}= -\oint_{\partial \Omega \left(\bm{r}_c, a_2, b_2 \right)} u(\bm{r}) \bm{w}^{(a)}_2 \cdot \bm{e}_n(\bm{r}) d \ell, \label{G2adefinition}
\end{align}
\begin{align}
    G^{(2b)} = -\oint_{\partial \Omega \left(\bm{r}_c, a_2, b_2 \right)} u(\bm{r}) \bm{w}^{(b)}_2 \cdot \bm{e}_n(\bm{r}) d \ell. \label{G2bdefinition}
\end{align}
We first calculate the steady-state concentration field in the case that the droplet is moving at a constant velocity $v \bm{e}_x$ and constant deformation amplitudes $a_2$ and $b_2$ as shown in Appendix~\ref{App_D}.
Then, we plug it into Eqs.~\eqref{Fdefinition}, \eqref{G2adefinition}, and \eqref{G2bdefinition} to obtain the explicit forms for $\bm{F}$, $G^{(2a)}$, and $G^{(2b)}$ as
\begin{align}
\bm{F} =& \left[ \left( F^{(x)}_1 + \tilde{F}^{(x)}_1 a_2 \right) v +\left( F^{(x)}_3 +  \tilde{F}^{(x)}_3 a_2  \right) v^3 \right] \bm{e}_x \nonumber \\
&+ \left[ \tilde{F}^{(y)}_1 b_2 v + \tilde{F}^{(y)}_3 b_2 v^3 \right] \bm{e}_y,
\end{align}
\begin{align}
G^{(2a)} =  G_2^{(2a)} v^2 + \tilde{G}^{(2a)}_0 a_2 + \tilde{G}_2^{(2a)} a_2 v^2,
\end{align}
\begin{align}
G^{(2b)} = \tilde{G}^{(2b)}_0 b_2 + \tilde{G}^{(2b)}_2 b_2 v^2,
\end{align}
where
\begin{align}
F^{(x)}_1 = \frac{\pi R^4}{4A} \left[ \mathcal{I}_0\left(R\right) \mathcal{K}_0\left(R \right) -  \mathcal{I}_2\left(R \right) \mathcal{K}_2\left(R \right) \right], \label{F1}
\end{align}
\begin{align}
F^{(x)}_3 =& \frac{\pi R^6}{32 A} \left[ \mathcal{I}_0(R) \mathcal{K}_0(R) - \frac{2}{R^2} \mathcal{I}_1(R) \mathcal{K}_1(R)  \right. \nonumber \\
& \left. \qquad  - \mathcal{I}_2(R) \mathcal{K}_2(R) \right],
\end{align}
\begin{align}
\tilde{F}^{(x)}_1 = \tilde{F}^{(y)}_1 = - \frac{\pi R^4}{2A} \left[ \mathcal{I}_1\left(R\right) \mathcal{K}_1\left(R \right)-  \mathcal{I}_2\left(R \right) \mathcal{K}_2\left(R \right) \right],
\end{align}
\begin{align}
\tilde{F}^{(x)}_3 =2\tilde{F}^{(y)}_3 = & \frac{\pi R^6}{48A} \left[ 3 \mathcal{I}_0\left(R\right) \mathcal{K}_0\left(R \right) - 4 \mathcal{I}_1\left(R\right) \mathcal{K}_1\left(R \right) \right. \nonumber \\
& \left. \qquad + \mathcal{I}_2\left(R \right) \mathcal{K}_2\left(R \right) \right],
\end{align}
\begin{align}
G^{(2a)}_2 =& \frac{\pi R^6}{64A}\left[ -2 \mathcal{I}_0(R) \mathcal{K}_0(R) +  \mathcal{I}_1(R) \mathcal{K}_1(R) \right. \nonumber \\
& \left. \qquad + 2 \mathcal{I}_2(R) \mathcal{K}_2(R) -  \mathcal{I}_3(R) \mathcal{K}_3(R)\right],
\end{align}
\begin{align}
\tilde{G}^{(2a)}_0 = \tilde{G}^{(2b)}_0 = \frac{\pi R^4}{A} \left[ \mathcal{I}_1\left(R\right) \mathcal{K}_1\left(R \right) -  \mathcal{I}_2\left(R \right) \mathcal{K}_2\left(R \right) \right], \label{G2a0}
\end{align}
\begin{align}
\tilde{G}^{(2a)}_2 = \tilde{G}^{(2b)}_2 =& \frac{ \pi R^6}{32 A} \left[ -4 \mathcal{I}_0(R) \mathcal{K}_0(R) + 7 \mathcal{I}_1(R) \mathcal{K}_1(R) \right. \nonumber \\
& \left. \qquad - 4 \mathcal{I}_2(R) \mathcal{K}_2(R) + \mathcal{I}_3(R) \mathcal{K}_3(R) \right]. \label{tildeG2a}
\end{align}
It is notable that $F^{(x)}_1 > 0$，$F^{(x)}_3 < 0$，$\tilde{F}^{(x)}_1 = \tilde{F}^{(y)}_1 < 0$, $G_2^{(2a)} < 0$, $\tilde{G}_0^{(2a)} = \tilde{G}_0^{(2b)} >0$ for all $R>0$. The other coefficients $\tilde{F}_3^{(x)}$, $\tilde{F}_3^{(y)}$, $\tilde{G}_2^{(2a)}$, and $\tilde{G}_2^{(2b)}$ change their signs depending on $R$.

In order to perform analyses based on dynamical systems, we introduce an inertia term to the equation of translational motion in Eq.~\eqref{drcdt_dl} as
\begin{align}
    m \frac{d \bm{v}_c}{dt} = -\eta_t \bm{v}_c  + \bm{F}.
\end{align}
The addition of the inertia term $m \, d\bm{v}_c/dt$ with effective mass $m$ is justified by the following two reasons. The first is that the particle should have a mass even if it is small, and the second is that the inertia term naturally appears by the expansion of the concentration field with respect to the time derivative of the velocity~\cite{Koyano_1D_2016,Koyano_PRE_rotor2017}.

\section{Analysis of ODE model \label{sec:ODE}}

In the previous section, the analysis is made under the constraint that the velocity is in the positive $x$-axis direction. Taking rotational symmetry of the system into consideration, the evolution equation is given as
\begin{align}
m \frac{d\bm{v}_c}{dt} =& (-\eta_t + f_1) \bm{v}_c - \tilde{f}_1 S \bm{v}_c - f_3 \left| \bm{v}_c\right|^2 \bm{v}_c \nonumber \\ &+ \tilde{f}_3 \left\{ \left| \bm{v}_c\right|^2 S \bm{v}_c + \left(\bm{v}_c \cdot S \bm{v}_c \right) \bm{v}_c \right\}, \label{ODE1}
\end{align}
\begin{align}
\eta_2 \frac{d S}{dt} = \left( -\kappa_2 + \tilde{g}_0 \right) S - g_2 \left( 2 \bm{v}_c \otimes \bm{v}_c - \left| \bm{v}_c\right|^2 I \right) - \tilde{g}_2 \left|\bm{v}_c\right|^2 S. \label{ODE2}
\end{align}
Here, $I$ is an identity tensor and $S$ is a traceless tensor that represents the second-order deformation, which is connected with $a_2$ and $b_2$~\cite{Ohta_Ohkuma2009,Tarama_2016,Ohta_JPSJ2017,hiraiwa2010} as
\begin{align}
S_{xx} = -S_{yy} = a_2,
\end{align}
\begin{align}
S_{xy} = S_{yx} = b_2.
\end{align}
The coefficients in Eqs.~\eqref{ODE1} and \eqref{ODE2} are given by Eqs.~\eqref{F1} to \eqref{tildeG2a} as $f_1 = F^{(x)}_1$, $\tilde{f}_1= -\tilde{F}^{(x)}_1 = -\tilde{F}^{(y)}_1 $, $f_3 = -F^{(x)}_3$, $\tilde{f}_3= \tilde{F}_3^{(x)}/2 = \tilde{F}_3^{(y)}$, $\tilde{g}_0 = \tilde{G}_0^{(2a)} = \tilde{G}_0^{(2b)}$, $g_2 = -G^{(2a)}_2$, and $\tilde{g}_2 = -\tilde{G}_2^{(2a)} = -\tilde{G}_2^{(2b)}$. These coefficients have positive values ($f_1 = 0.07812$, $\tilde{f}_1 = 0.05980$, $f_3 \simeq 0.01150$, $\tilde{f}_3 \simeq 0.09292$, $\tilde{g}_0 \simeq 0.1196$, $g_2 \simeq 0.006909$, $\tilde{g}_2 \simeq 0.01487$) when $R=1$, which corresponds to the numerical simulation in Section~\ref{sec:simulation}.

Considering the terms up to the third order of small parameters $\bm{v}_c$ and $S$, we obtain the ordinary differential equation (ODE) model:
\begin{align}
m\frac{d\bm{v}_c}{dt} = \left(-\eta_t + f_1\right) \bm{v}_c - \tilde{f}_1 S \bm{v}_c - f_3\left| \bm{v}_c\right|^2 \bm{v}_c, \label{eq58}
\end{align}
\begin{align}
\eta_2 \frac{d S}{dt} =&  \left( -\kappa_2 + \tilde{g}_0 \right) S - g_2 \left( 2 \bm{v}_c \otimes \bm{v}_c - \left| \bm{v}_c\right|^2 I \right) \nonumber \\ &- \tilde{g}_2 \left|\bm{v}_c\right|^2 S - h_0 S^3. \label{eq59}
\end{align}
It should be noted that here we phenomenologically add the term $-h_0 S^3$ in order to realize stable deformation of the droplet, though the term proportional to $S^3$ does not appear from the expansion in the discussion above since we only consider the first order of the deformation amplitude. Actually, the value of $h_0$ in the case of $R=1$, corresponding to the numerical simulation in Section~\ref{sec:simulation}, can be determined by the fitting and it is given as $h_0 \simeq 0.1436$ (see the details in Appendix~\ref{app:h0}).

\begin{figure*}
    \centering
    \includegraphics{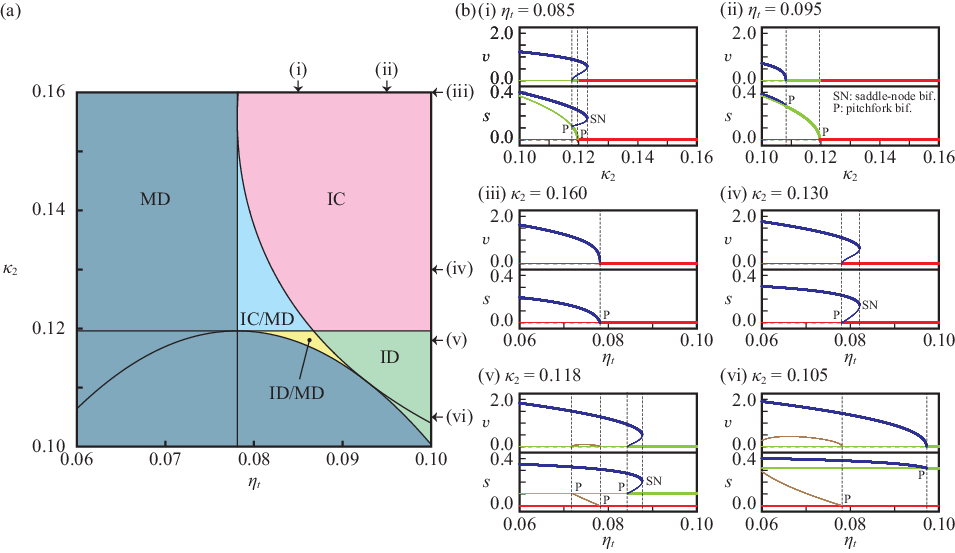}
    \caption{Analysis of the ODE model in Eqs.~\eqref{eq58} and \eqref{eq59}. (a) Phase diagram obtained by the ODE model. The boundary curves are given as $\eta_t = f_1$, $\kappa_2 = \tilde{g}_0$, Eq.~\eqref{funcboundary} with Eq.~\eqref{Hs}, and Eq.~\eqref{idmd}. IC, ID, and MD indicate an immobile circular droplet, an immobile deformed droplet, and a mobile deformed droplet moving in the minor-axis direction, respectively. IC/MD and ID/MD denote the bistable states of IC and MD droplets and ID and MD droplets, respectively. (b) One-dimensional bifurcation diagrams with constant $\eta_t$ or $\kappa_2$, which are indicated as arrows at the top and right sides of the plot in panel (a). Speed $v$ and deformation $s$ are plotted. (i,ii) $\kappa_2$ is varied at (i) $\eta_t = 0.085$ and (ii) $\eta_t = 0.095$. (iii-vi) $\eta_t$ is varied at (iii) $\kappa_2 = 0.160$, (iv) $\kappa_2 = 0.130$, (v) $\kappa_2 = 0.118$, and (vi) $\kappa_2 = 0.105$. The red, green, dark blue, and brown lines represent immobile circular (IC), immobile deformed (ID), mobile deformed (MD) droplets moving in the minor-axis direction, and MD droplets moving in the major-axis direction. The thick and thin lines indicate the stable and unstable solutions, respectively. The characters ``P'' and ``SN'' mean the pitchfork and saddle-node bifurcations. The vertical dotted lines are drawn to indicate the transition points and correspondence between $v$ and $s$.}
    \label{fig3}
\end{figure*}

For the stability analysis, we consider in the polar coordinates as
\begin{align}
\bm{v}_c = v \left( \cos \phi \, \bm{e}_x + \sin \phi \, \bm{e}_y \right),
\end{align}
\begin{align}
S = s \begin{pmatrix} \cos 2\psi & \sin 2\psi  \\ \sin 2\psi & - \cos2\psi \end{pmatrix} = \begin{pmatrix} a_2 & b_2 \\ b_2 & - a_2 \end{pmatrix}. 
\end{align}
It should be noted that the angle $\psi$ indicates the elongated direction of the 2-mode deformation from a circle and the deformation amplitude is denoted by $s$.
Here, we calculate $S^3$ as
\begin{align}
(S^3)_{\alpha \beta} = S_{\alpha \gamma} S_{\gamma \delta} S_{\delta \beta} = s^2 \delta_{\alpha \delta} S_{\delta \beta} = s^2 S_{\alpha \beta}.
\end{align}
Therefore, we finally obtain the ODE model in the polar coordinates as
\begin{align}
m\frac{dv}{dt} = \left(-\eta_t + f_1\right) v - \tilde{f}_1 s v \cos 2(\psi  - \phi) - f_3 v^3, \label{eq_v}
\end{align}
\begin{align}
m \frac{d\phi}{dt} = - \tilde{f}_1 s \sin 2(\psi - \phi), \label{eq_phi}
\end{align}
\begin{align}
\eta_2 \frac{ds}{dt} = \left(-\kappa_2 + \tilde{g}_0 \right) s - g_2 v^2 \cos 2(\psi - \phi) - \tilde{g}_2 v^2 s - h_0 s^3, \label{eq_s}
\end{align}
\begin{align}
\eta_2 \frac{d\psi}{dt} = \frac{1}{2} g_2 \frac{v^2}{s} \sin 2(\psi - \phi). \label{eq_psi}
\end{align}
Based on the ODE model, we perform the analyses of the solutions and their stability. The details are shown in Appendix~\ref{app:detail}.

We adopt the coefficient values that correspond to the numerical simulation ($R=1$) by the partial differential equation (PDE) model in Section~\ref{sec:simulation} and obtain the phase diagram and the bifurcation diagrams shown in Fig.~\ref{fig3}. These diagrams well reproduce the ones obtained from the direct numerical simulation using the PDE model shown in Fig.~\ref{fig2}. The amplitudes of $v$ and $s$ in the bifurcation diagrams also meet between the PDE and ODE models. The quantitative correspondence is closer in the parameter region near the bifurcation point from the IC droplet, i.e., $\eta_t = f_1$ and $\kappa_2 = \tilde{g}_0$. Far from the bifurcation point, the quantitative correspondence becomes worse, but the qualitative one is maintained.

From the bifurcation diagrams in Fig.~\ref{fig3}(b), we can determine the bifurcation types between the different modes. An IC droplet becomes unstable and a minor-axis directed moving elliptic droplet [a mobile deformed (MD) droplet] becomes stable with decreasing $\eta_t$. This transition is classified into a supercritical drift-pitchfork bifurcation for $\kappa_2 \gtrsim 0.156$ and a subcritical one for $\kappa_2 \lesssim 0.156$. It should be noted that this drift-pitchfork bifurcation is symmetric in the two dimension. With a decrease in $\kappa_2$, an IC droplet becomes unstable and an ID droplet becomes stable through a supercritical pitchfork bifurcation. It should be noted that this bifurcation is also symmetric in the two dimension. With a further decrease in $\kappa_2$, an ID droplet becomes unstable and an MD droplet which moves in the minor-axis direction becomes stable. This transition is classified into a supercritical drift-pitchfork bifurcation for $\eta_t \gtrsim 0.0925$ and a subcritical one for $0.0781 < \eta_t \lesssim 0.0925$. For the MD droplet, only the minor-axis directed motion is stable. For $\kappa_2 \gtrsim 0.156$, it becomes unstable and an IC droplet becomes stable with an increase in $\eta_t$ via a supercritical drift-pitchfork bifurcation. For $0.114 \lesssim \kappa_2 \lesssim 0.156$, the MD droplet becomes unstable via a saddle-node bifurcation. The unstable solution of the MD droplet connects the stable MD droplet and the stable IC or ID droplet. For $\kappa_2 \lesssim 0.114$, the MD droplet becomes unstable and the ID droplet becomes stable via a supercritical drift-pitchfork bifurcation. This suggests the degenerate pitchfork bifurcation among the solutions corresponding to the IC, ID and MD droplets. In addition, the bifurcation analysis reveals that the unstable solution corresponding to the deformed droplet moving in the major-axis direction, which connects the solutions for IC and ID droplets, is bifurcated from them via supercritical drift-pitchfork bifurcations. The simulation results using the PDE model can be explained by the results of the bifurcation analysis without any contradiction.

\section{Discussion and Conclusion}\label{sec:discussion}

In the present paper, we constructed a model for a deformable self-propelled droplet that exhibits self-diffusiophoresis and performed numerical simulations and theoretical analyses to discuss the relation between motion and deformation using the Fourier expansion for the deformation.
The analyses based on the Fourier mode expansion were also reported in previous studies. Ohta and Ohkuma proposed a mathematical model in which the velocity of the self-propelled particle and the amplitude of 2-mode deformation are considered~\cite{Ohta_Ohkuma2009,Ohta_JPSJ2017}. They propose the model only by considering the symmetric properties of the system. In contrast, we derived the coefficients appearing in the Ohta-Ohkuma model by the reduction from a model including a concentration profile. The obtained results suggest both spontaneous translational motion and spontaneous deformation are possible. In the paper by Ohta and Ohkuma~\cite{Ohta_Ohkuma2009}, they limited their analysis to the case of no spontaneous deformation, i.e., $-\kappa_2 + \tilde{g}_0 < 0$. In addition, the combination of the signs of the coupling terms between the velocity and deformation was $\tilde{f}_1 g_2 < 0$, which is different from our results. As the continuing study of \cite{Ohta_Ohkuma2009}, Hiraiwa et al.~\cite{hiraiwa2010} and Tarama et al.~\cite{Tarama_2016} proposed the models with the spontaneous deformation, in which the cubic term of $S$ is included. In their models, they also limited their analysis to the case of the coupling terms as $\tilde{f}_1 g_2 < 0$. Therefore, the bifurcation structure with the bistable states between IC and MD droplets and between ID and MD droplets is unique to our model. Moreover, we can show that an ID droplet can stably exist in the parameter region derived from the PDE model.

Our problem has parallels with the dynamics of a localized pulse in two-dimensional reaction-diffusion systems.
There have been several studies in which the coupling between the pulse shape and motion is concerned. 
Krischer and Mikhailov reported a traveling pulse in a two-dimensional reaction-diffusion system with volume conservation using numerical simulation. They exhibited a characteristic shape of a traveling solution~\cite{Krischer_Mikhailov_1994}. Teramoto et al. studied the reaction-diffusion system with one activator and two inhibitors. They analyzed the bifurcation structure based on the center manifold theorem. They clarified the bifurcation structure including the bifurcation from a circular-shaped standing pulse into a deformed traveling pulse solution~\cite{Teramoto_Nishiura_2009}.
Ohta et al. analyzed the reaction-diffusion equation with volume conservation and discussed the dynamics by expanding with Fourier series~\cite{Ohta_Ohkuma_Shitara_2009}. They also expanded their study into a three-dimensional system~\cite{Shitara_Hiraiwa_Ohta_2011}.
Our approach is similar to their work in that the shape is described as the Fourier expansion in polar coordinates. Since our reaction-diffusion equation is linear, the analysis is simpler and the analytical approach can be more easily performed.

Recently, some of the authors published a paper on the phase-field model of a self-propelled droplet almost in the same mechanism, in which a droplet shape is described by a phase field and the evolution equation was derived by the variational principle of free energy. There, the relation between the time evolution of the phase field and that of the interface was discussed~\cite{Nagayama2023_SciRep}. This work can be considered to be the case where the droplet shape is described as a function of the angle and the function is expanded as the Fourier series. Thanks to this simplification, we can discuss the stability of motion in our present model, though we could only perform numerical simulations to obtain the droplet motion in our previous model. We hope the universal understanding of a self-propelled deformed droplet will be obtained by the comparison between the previous phase-field model and the present model using the Fourier expansion.  


We constructed a mathematical model for a deformable diffusiophoretic self-propelled droplet, which moves by the surface tension gradient originating from the concentration field of the chemicals emitted from the droplet. By defining the free energy of the system including the surface and line energies, and calculating the variation of it, we obtained the time-evolution equations for the translational motion and deformation. In the case only including the 2-mode deformation, we performed numerical simulations and theoretical analyses, and obtained that an immobile deformed droplet or a mobile deformed droplet that moves in its minor-axis direction becomes stable through a supercritical and subcritical pitchfork bifurcation from an immobile circular droplet, respectively.

\begin{acknowledgments}
This work was supported by JSPS KAKENHI Grants Nos.~JP22K03428, JP23K20815, JP23H04936, JP24K06978, JP24K22311, JP24K16981, JP25K00918,  and the Cooperative Research Program of ``Network Joint Research Center for Materials and Devices'' (Nos.~20254003, 20251011).
This work was also supported by JSPS and PAN under the Japan-Poland Research Cooperative Program (No.~JPJSBP120204602) and the JSPS Core-to-Core Program ``Advanced core-to-core network for the physics of self-organizing active matter (JPJSCCA20230002)''.
This work was also supported by the Research Institute for Mathematical Sciences, an International Joint Usage/Research Center located in Kyoto University and by MEXT Promotion of Distinctive Joint Usage/Research Center Support Program Grant Number JPMXP0724020292.
\end{acknowledgments}

\appendix

\onecolumngrid

\section{Detailed derivation of $\partial E/ \partial \bm{r}_c$, $\partial E/ \partial a_k$, and $\partial E/ \partial b_k$ in Eqs.~\eqref{dedrc}, \eqref{dedak}, and \eqref{dedbk} in the main text \label{App_C}}

Here, we show the detailed derivation of Eqs.~\eqref{dedrc}, \eqref{dedak}, and \eqref{dedbk} in the main text. We set 
$\bm{r}_c = x_c \bm{e}_x + y_c \bm{e}_y$, and first we calculate $\partial E_s/\partial x_c$ as
\begin{align}
\frac{\partial E_s}{\partial x_c} =& \lim_{\delta \to 0} \frac{\iint_{\mathbb{R}^2 \backslash\Omega(\bm{r}_c + \delta \bm{e}_x, \left\{ a_k \right\}, \left\{ b_k\right\})} \gamma(u(\bm{r})) dA - \iint_{\mathbb{R}^2 \backslash\Omega(\bm{r}_c, \left\{ a_k \right\}, \left\{ b_k\right\})} \gamma(u(\bm{r})) dA}{\delta} \nonumber \\
=& -\lim_{\delta \to 0} \frac{\iint_{\Omega(\bm{r}_c + \delta \bm{e}_x, \left\{ a_k \right\}, \left\{ b_k\right\})} \gamma(u(\bm{r})) dA - \iint_{\Omega(\bm{r}_c, \left\{ a_k \right\}, \left\{ b_k\right\})} \gamma(u(\bm{r})) dA}{\delta} \nonumber \\
=& -\lim_{\delta \to 0} \frac{\iint_{\Omega(\bm{r}_c, \left\{ a_k \right\}, \left\{ b_k\right\})} \gamma(u(\bm{r} + \delta \bm{e}_x)) dA - \iint_{\Omega(\bm{r}_c, \left\{ a_k \right\}, \left\{ b_k\right\})} \gamma(u(\bm{r})) dA}{\delta} \nonumber \\
=& - \lim_{\delta \to 0} \iint_{\Omega(\bm{r}_c, \left\{ a_k \right\}, \left\{ b_k\right\})} \frac{\gamma(u(\bm{r} + \delta \bm{e}_x)) - \gamma(u(\bm{r}))}{\delta} dA \nonumber \\
=& - \iint_{\Omega(\bm{r}_c, \left\{ a_k \right\}, \left\{ b_k\right\})} \frac{ \partial \gamma(u(\bm{r}))}{\partial x} dA.
\end{align}
Here, we invert the integrated region from the first to the second lines. Then, we replace $\bm{r}$ with $\bm{r} + \delta \bm{e}_x$ only for the first term of the numerator and adopt the definition of the derivative.   
In the same manner, we obtain
\begin{align}
\frac{\partial E_s}{\partial y_c} 
=& - \iint_{\Omega(\bm{r}_c, \left\{ a_k \right\}, \left\{ b_k\right\})} \frac{ \partial \gamma(u(\bm{r}))}{\partial y} dA.
\end{align}
and therefore we get
\begin{align}
\frac{\partial E_s}{\partial \bm{r}_c} =& - \iint_{\Omega(\bm{r}_c, \left\{ a_k \right\}, \left\{ b_k\right\})} \nabla \gamma(u(\bm{r})) dA.
\end{align}
Using the Gauss' divergence law,
\begin{align}
\iint_{\Omega(\bm{r}_c, \left\{ a_k \right\}, \left\{ b_k\right\})} \frac{ \partial \gamma(u(\bm{r}))}{\partial x} dA &= \iint_{\Omega(\bm{r}_c, \left\{ a_k \right\}, \left\{ b_k\right\})} \bm{e}_x \cdot \nabla \gamma(u(\bm{r})) dA \nonumber \\
&= \iint_{\Omega(\bm{r}_c, \left\{ a_k \right\}, \left\{ b_k\right\})} \nabla \cdot  (\gamma(u(\bm{r})) \bm{e}_x ) dA \nonumber \\
&= \oint_{\partial \Omega(\bm{r}_c, \left\{ a_k \right\}, \left\{ b_k\right\})} \gamma(u(\bm{r})) \bm{e}_x \cdot \bm{e}_n(\bm{r}) d\ell.
\end{align}
Here, we use
\begin{align}
    \nabla \cdot \left( \gamma(u(\bm{r})) \bm{e}_x \right) = \nabla \gamma(u(\bm{r})) \cdot \bm{e}_x + \gamma(u(\bm{r})) \nabla \cdot \bm{e}_x = \nabla \gamma(u(\bm{r})) \cdot \bm{e}_x.  
\end{align}
In the same manner, we obtain
\begin{align}
\iint_{\Omega(\bm{r}_c, \left\{ a_k \right\}, \left\{ b_k\right\})} \frac{ \partial \gamma(u(\bm{r}))}{\partial y} dA &= \oint_{\partial \Omega(\bm{r}_c, \left\{ a_k \right\}, \left\{ b_k\right\})} \gamma(u(\bm{r})) \bm{e}_y \cdot \bm{e}_n(\bm{r}) d\ell.
\end{align}
Therefore,
\begin{align}
\iint_{\Omega(\bm{r}_c, \left\{ a_k \right\}, \left\{ b_k\right\})}  \nabla \gamma(u(\bm{r})) dA &= \oint_{\partial \Omega(\bm{r}_c, \left\{ a_k \right\}, \left\{ b_k\right\})} \gamma(u(\bm{r})) \left[ (\bm{e}_x \cdot \bm{e}_n(\bm{r})) \bm{e}_x + (\bm{e}_y \cdot \bm{e}_n(\bm{r})) \bm{e}_y \right] d\ell \nonumber \\
&= \oint_{\partial \Omega(\bm{r}_c, \left\{ a_k \right\}, \left\{ b_k\right\})} \gamma(u(\bm{r})) \bm{e}_n(\bm{r}) d\ell,
\end{align}
and we obtain Eq.~\eqref{dedrc} in the main text.

Next, we calculate $\partial E_s/\partial a_k$. We define
$\left\{ e_n^{(k)} \right\}$ as
\begin{align}
e_n^{(k)} = \delta_{nk},
\end{align}
where $\delta_{ij}$ is the Kronecker's delta. Using this, we obtain
\begin{align}
\frac{\partial E_s}{\partial a_k} =& \lim_{\delta \to 0} \frac{\iint_{\mathbb{R}^2 \backslash\Omega \left(\bm{r}_c, \left\{ a_k \right\} + \delta \left\{ e_n^{(k)} \right\}, \left\{ b_k \right\}\right)} \gamma(u(\bm{r})) dA - \iint_{\mathbb{R}^2 \backslash\Omega\left(\bm{r}_c, \left\{ a_k\right\}, \left\{ b_k\right\}\right)} \gamma(u(\bm{r})) dA}{\delta} \nonumber \\
=& -\lim_{\delta \to 0} \frac{\iint_{\Omega \left(\bm{r}_c, \left\{ a_k \right\} + \delta \left\{ e_n^{(k)} \right\}, \left\{ b_k \right\}\right)} \gamma(u(\bm{r})) dA - \iint_{\Omega \left(\bm{r}_c, \left\{ a_k\right\}, \left\{ b_k\right\}\right)} \gamma(u(\bm{r})) dA}{\delta} \nonumber \\
=&  -\lim_{\delta \to 0} \frac{\iint_{\Omega \left(\bm{r}_c, \left\{ a_k \right\}, \left\{ b_k \right\}\right)} \gamma(u(\bm{r} + \delta \bm{w}^{(a)}_k)) dA - \iint_{\Omega \left(\bm{r}_c, \left\{ a_k\right\}, \left\{ b_k\right\}\right)} \gamma(u(\bm{r})) dA}{\delta} \nonumber \\
=&  -\lim_{\delta \to 0} \frac{1}{\delta} \iint_{\Omega \left(\bm{r}_c, \left\{ a_k \right\}, \left\{ b_k \right\}\right)} \left[ \gamma(u(\bm{r} + \delta \bm{w}^{(a)}_k)) - \gamma(u(\bm{r}))\right] dA \nonumber \\
=&  -\lim_{\delta \to 0} \frac{1}{\delta} \iint_{\Omega \left(\bm{r}_c, \left\{ a_k \right\}, \left\{ b_k \right\}\right)} \left[\delta \bm{w}^{(a)}_k \cdot \nabla \gamma(u(\bm{r})) + \mathcal{O}(\delta^2)\right] dA \nonumber \\
=& - \iint_{\Omega \left(\bm{r}_c, \left\{ a_k \right\} , \left\{ b_k \right\} \right)} \bm{w}^{(a)}_k \cdot \nabla \gamma(u(\bm{r})) dA. 
\end{align}
In the calculation, we invert the integrated region from the first to the second lines. From the second to third lines, we replace $\bm{r}$ with $\bm{r} + \delta \bm{w}_k^{(a)}$, where $\bm{w}_k^{(a)}$ is given in Eq.~\eqref{wka} in the main text. The Jacobian matrix regarding this transform is $1 + \mathcal{O}(\epsilon^2)$.
The conditions required for the vector field $\bm{w}_k^{(a)}$ are that the divergence is 0, i.e., $\nabla \cdot \bm{w}_k^{(a)} = 0$ and that it satisfies
\begin{align}
f(\theta,\{a_k\}+\delta\{e_n^{(k)}\},\{b_k\}) \bm{e}_r = f(\theta,\{a_k\},\{b_k\}) \bm{e}_r + \delta \bm{w}^{(a)}_k|_{\bm{r} = f(\theta,\{a_k\},\{b_k\}) \bm{e}_r}. 
\end{align}
The incompressible flow profile in Eq.~\eqref{wka} in the main text holds the above conditions.

Using the Gauss' divergence theorem, we obtain
\begin{align}
\iint_{\Omega \left(\bm{r}_c, \left\{ a_k \right\} , \left\{ b_k \right\} \right)} \bm{w}^{(a)}_k \cdot \nabla \gamma(\bm{r}) d A =& \iint_{\Omega \left(\bm{r}_c, \left\{ a_k \right\} , \left\{ b_k \right\} \right)} \nabla \cdot \left(\gamma(u(\bm{r})) \bm{w}^{(a)}_k \right) dA
 \nonumber \\
=& \oint_{\partial \Omega \left(\bm{r}_c, \left\{ a_k \right\} , \left\{ b_k \right\} \right)} \gamma(u(\bm{r})) \bm{w}^{(a)}_k \cdot \bm{e}_n(\bm{r}) d \ell.
\end{align}
Here, we use 
\begin{align}
    \nabla \cdot \left( \gamma(u(\bm{r})) \bm{w}^{(a)}_k \right) = \nabla \gamma(u(\bm{r})) \cdot \bm{w}^{(a)}_k + \gamma(u(\bm{r})) \nabla \cdot \bm{w}^{(a)}_k = \nabla \gamma(u(\bm{r})) \cdot \bm{w}^{(a)}_k.  
\end{align}
Therefore, we obtain
\begin{align}
\frac{\partial E_s}{\partial a_k} =& - \iint_{\Omega \left(\bm{r}_c, \left\{ a_k \right\} , \left\{ b_k \right\} \right)}\bm{w}^{(a)}_k \cdot \nabla \gamma(u(\bm{r})) d A \nonumber \\
=& - \oint_{\partial \Omega \left(\bm{r}_c, \left\{ a_k \right\} , \left\{ b_k \right\} \right)} \gamma(u(\bm{r})) \bm{w}^{(a)}_k \cdot \bm{e}_n(\bm{r}) d \ell.
\end{align}
\begin{align}
    \nabla \cdot \left( \gamma(u(\bm{r}) \bm{e}_x \right) = \nabla \gamma(u(\bm{r})) \cdot \bm{e}_x + \gamma(u(\bm{r})) \nabla \cdot \bm{e}_x = \nabla \gamma(u(\bm{r})) \cdot \bm{e}_x.  
\end{align}
In the same manner, we obtain
\begin{align}
\frac{\partial E_s}{\partial b_k} =& - \iint_{\Omega \left(\bm{r}_c, \left\{ a_k \right\} , \left\{ b_k \right\} \right)} \bm{w}^{(b)}_k \cdot \nabla \gamma (u(\bm{r})) d A \nonumber \\
=& - \oint_{\partial \Omega \left(\bm{r}_c, \left\{ a_k \right\} , \left\{ b_k \right\} \right)} \gamma (u(\bm{r})) \bm{w}^{(b)}_k \cdot \bm{e}_n(\bm{r}) d \ell ,
\end{align}
and thus we get Eqs.~\eqref{dedak} and \eqref{dedbk} in the main text.

\section{Detailed derivation of the steady-state concentration \label{App_D}}

Here, we show the steady-state concentration $u_{s}$ when the droplet is moving at a constant velocity $\bm{v}_c$ and constant deformation amplitudes $a_2$ and $b_2$. For simplicity, we assume $\bm{v}_c = v\bm{e}_x$. The concentration field should hold
\begin{align}
    -v \frac{\partial u_s}{\partial x} = \frac{\partial^2 u_s}{\partial x^2} + \frac{\partial^2 u_s}{\partial y^2} - u_s + S(\bm{r}; \bm{r}_c, a_2, b_2), \label{eqS13}
\end{align}
with
\begin{align}
S(\bm{r}; \bm{r}_c, a_2, b_2) =  \left\{ \begin{array}{ll} 1 / A, & \bm{r} \in \Omega(\bm{r}_c, a_2, b_2), \\ 0, & \bm{r} \notin \Omega(\bm{r}_c, a_2, b_2). \end{array} \right. \label{eqS14}
\end{align}
and
\begin{align}
\Omega(\bm{r}_c,  a_2, b_2) = \left\{  \bm{r} =  \bm{r}_c + r \bm{e}_r(\theta) \middle| r \leq R \left(1 + a_2 \cos2\theta + b_2 \sin 2\theta \right)\right\}. \label{Omega0_2}
\end{align}
We also need to consider the continuity condition at the droplet boundary. The results are already given in our previous papers~\cite{Kitahata_Iida_2013,iida2014theoretical,Kitahata_JPSJ_2020,Kitahata_Frontier2022} on a self-propelled solid particle with a deformed shape from a circle. Here, we set $a_2 = \epsilon \tilde{a}_2$ and $b_2 = \epsilon \tilde{b}_2$ to clearly show the dependence of the small parameter $\epsilon$. Then, we obtain
\begin{align}
    u_s = \left\{ \begin{array}{ll} u_s^{(\mathrm{i})} = u^{(\mathrm{i})}_0 + u^{(\mathrm{i})}_1 v + u^{(\mathrm{i})}_2 v^2 + u^{(\mathrm{i})}_3 v^3 + \epsilon \left(\tilde{u}^{(\mathrm{i})}_0 + \tilde{u}^{(\mathrm{i})}_1 v + \tilde{u}^{(\mathrm{i})}_2 v^2 + \tilde{u}^{(\mathrm{i})}_3 v^3\right) + \mathcal{O}(v^4, \epsilon^2) , & \bm{r} \in \Omega(\bm{r}_c, a_2, b_2), \\ u_s^{(\mathrm{o})} = u^{(\mathrm{o})}_0 + u^{(\mathrm{o})}_1 v + u^{(\mathrm{o})}_2 v^2 + u^{(\mathrm{o})}_3 v^3 + \epsilon \left(\tilde{u}^{(\mathrm{o})}_0 + \tilde{u}^{(\mathrm{o})}_1 v + \tilde{u}^{(\mathrm{o})}_2 v^2 + \tilde{u}^{(\mathrm{o})}_3 v^3\right) + \mathcal{O}(v^4, \epsilon^2), & \bm{r} \notin \Omega(\bm{r}_c, a_2, b_2), \end{array} \right. \label{us}
\end{align}
where the explicit forms are given in the polar coordinates as
\begin{align}
u_0^{(\mathrm{i})} = \frac{1}{A} \left[ 1 - R \mathcal{K}_1\left(R\right) \mathcal{I}_0 \left(r \right) \right],
\end{align}
\begin{align}
u_{0}^{(\mathrm{o})} = \frac{1}{A}  \mathcal{I}_1 \left(R \right) \mathcal{K}_0 \left(r \right),
\end{align}
\begin{align}
u_{1}^{(\mathrm{i})} = \frac{R}{2A} \left[ r \mathcal{K}_1\left(R\right) \mathcal{I}_0\left(r \right)- R \mathcal{K}_2 \left(R \right) \mathcal{I}_1\left(r \right) \right] \cos\theta ,
\end{align}
\begin{align}
u_{1}^{(\mathrm{o})} = \frac{R}{2A} \left[ -r \mathcal{I}_1\left(R\right) \mathcal{K}_0\left(r \right)+ R \mathcal{I}_2 \left(R \right) \mathcal{K}_1\left(r \right) \right] \cos\theta ,
\end{align}
\begin{align}
u_{2}^{(\mathrm{i})} =& \frac{R^2}{32A} \left[ r^2\left( \mathcal{K}_0(R) \mathcal{I}_0(r) - \mathcal{K}_2(R) \mathcal{I}_2(r) \right) - 2 \left(R \mathcal{K}_1(R) \mathcal{I}_0(r) - r \mathcal{K}_0(R) \mathcal{I}_1(r) \right) ,\right] \nonumber \\
&+ \frac{R^2}{64A} \left[ 2 r^2 \left( \mathcal{K}_0(R) \mathcal{I}_0(r) - \mathcal{K}_2(R) \mathcal{I}_2(r) \right) - R r \left( \mathcal{K}_1(R) \mathcal{I}_1(r) - \mathcal{K}_3(R) \mathcal{I}_3(r) \right) \right] \cos 2 \theta ,
\end{align}
\begin{align}
u_{2}^{(\mathrm{o})} =& \frac{R^2}{32A} \left[ r^2\left( \mathcal{I}_0(R) \mathcal{K}_0(r) - \mathcal{I}_2(R) \mathcal{K}_2(r) \right) + 2 \left(R \mathcal{I}_1(R) \mathcal{K}_0(r) - r \mathcal{I}_0(R) \mathcal{K}_1(r) \right) \right] \nonumber \\
&+ \frac{R^2}{64A} \left[ 2 r^2 \left( \mathcal{I}_0(R) \mathcal{K}_0(r) - \mathcal{I}_2(R) \mathcal{K}_2(r) \right) - R r \left( \mathcal{I}_1(R) \mathcal{K}_1(r) - \mathcal{I}_3(R) \mathcal{K}_3(r) \right) \right] \cos 2 \theta,
\end{align}
\begin{align}
u_{3}^{(\mathrm{i})} =& \frac{R^2}{128A} \left[ r (R^2 + r^2) \left( \mathcal{K}_2(R) \mathcal{I}_2(r) - \mathcal{K}_0(R) \mathcal{I}_0(r)\right) + 4 r \left(R \mathcal{K}_1(R) \mathcal{I}_0(r) - r \mathcal{K}_0(R) \mathcal{I}_1(r)\right) \right] \cos \theta \nonumber \\
& + \frac{R^2}{1152A}\left[ -3 r^3 \left( \mathcal{K}_0(R)\mathcal{I}_0(r) - \mathcal{K}_2(R) \mathcal{I}_2(r)\right) + 3 r^2 R \left( \mathcal{K}_1(R) \mathcal{I}_1(r) - \mathcal{K}_3(R) \mathcal{I}_3(r) \right) \right. \nonumber \\
&\left. \qquad - R^2 r \left(\mathcal{K}_2(R) \mathcal{I}_2 (r) - \mathcal{K}_4(R) \mathcal{I}_4(r) \right) \right] \cos 3\theta,
\end{align}
\begin{align}
u_{3}^{(\mathrm{o})} =&\frac{R^2}{128A} \left[ r (R^2 + r^2) \left( \mathcal{I}_2(R) \mathcal{K}_2(r) - \mathcal{K}_0(R) \mathcal{K}_0(r)\right) - 4 r \left(R \mathcal{I}_1(R) \mathcal{K}_0(r) - r \mathcal{I}_0(R) \mathcal{K}_1(r)\right) \right] \cos \theta \nonumber \\
& + \frac{R^2}{1152A}\left[ -3 r^3 \left( \mathcal{I}_0(R)\mathcal{K}_0(r) - \mathcal{I}_2(R) \mathcal{K}_2(r)\right) + 3 r^2 R \left( \mathcal{I}_1(R) \mathcal{K}_1(r) - \mathcal{I}_3(R) \mathcal{K}_3(r) \right) \right. \nonumber \\
&\left. \qquad - R^2 r \left(\mathcal{I}_2(R) \mathcal{K}_2 (r) - \mathcal{I}_4(R) \mathcal{K}_4(r) \right) \right] \cos 3\theta,
\end{align}
\begin{align}
    \tilde{u}^{(i)}_0 =& \frac{1}{A} \left[ R^2 \mathcal{K}_2(R) \mathcal{I}_2(r) \right] \left( \tilde{a}_2 \cos 2 \theta + \tilde{b}_2 \sin 2\theta \right),
\end{align}
\begin{align}
    \tilde{u}^{(\mathrm{o})}_0 = \frac{1}{A} \left[R^2 \mathcal{I}_2(R) \mathcal{K}_2(r)\right] \left( \tilde{a}_2 \cos 2 \theta + \tilde{b}_2 \sin 2\theta \right),
\end{align}
\begin{align}
    \tilde{u}^{(\mathrm{i})}_1 =& \frac{R^2}{4A} \left[ -r \mathcal{K}_2(R) \mathcal{I}_2(r) + R \mathcal{K}_1(R) \mathcal{I}_1(r) \right] \left( \tilde{a}_2 \cos \theta + \tilde{b}_2 \sin\theta\right) \nonumber \\
    & + \frac{R^2}{4A} \left[- r \mathcal{K}_2(R) \mathcal{I}_2(r) + R \mathcal{K}_3(R) \mathcal{I}_3(r) \right] \left( \tilde{a}_2 \cos 3 \theta + \tilde{b}_2 \sin 3\theta \right),
\end{align}
\begin{align}
    \tilde{u}^{(\mathrm{o})}_1 =& \frac{R^2}{4A} \left[ -r \mathcal{I}_2(R) \mathcal{K}_2(r) + R \mathcal{I}_1(R) \mathcal{K}_1(r) \right] \left( \tilde{a}_2 \cos \theta + \tilde{b}_2 \sin \theta \right) \nonumber \\
    & + \frac{R^2}{4A} \left[- r \mathcal{I}_2(R) \mathcal{K}_2(r) + R \mathcal{I}_3(R) \mathcal{K}_3(r) \right] \left( \tilde{a}_2 \cos 3\theta + \tilde{b}_2 \sin 3\theta \right) ,
\end{align}
\begin{align}
 \tilde{u}^{(\mathrm{i})}_2 =& \frac{R^2}{32A} \left[ R^2 \mathcal{K}_0(R) \mathcal{I}_0(r) - 2 r R \mathcal{K}_1(R) \mathcal{I}_1(r) + r^2 \mathcal{K}_2(R) \mathcal{I}_2(r) \right] \tilde{a}_2 \nonumber \\
 &+ \frac{R^2}{16A} \left[ -\frac{3}{2} r R \mathcal{K}_1(R) \mathcal{I}_1(r) + (r^2 + R^2) \mathcal{K}_2(R) \mathcal{I}_2(r) - \frac{1}{2} r R \mathcal{K}_3(R) \mathcal{I}_3(r) \right] \left( \tilde{a}_2 \cos 2\theta + \tilde{b}_2 \sin 2\theta \right) \nonumber \\
 &+ \frac{R^2}{32A} \left[ r^2 \mathcal{K}_2(R) \mathcal{I}_2(r) - 2 r R \mathcal{K}_3(R) \mathcal{I}_3(r) + R^2 \mathcal{K}_4(R) \mathcal{I}_4(r) \right] \left( \tilde{a}_2 \cos 4\theta + \tilde{b}_2 \sin 4\theta \right),
\end{align}
\begin{align}
 \tilde{u}^{(\mathrm{o})}_2 =& \frac{R^2}{32A} \left[ R^2 \mathcal{I}_0(R) \mathcal{K}_0(r) - 2 r R \mathcal{I}_1(R) \mathcal{K}_1(r) + r^2 \mathcal{I}_2(R) \mathcal{K}_2(r) \right] \nonumber \\
 &+ \frac{R^2}{16A} \left[-\frac{3}{2} r R \mathcal{I}_1(R) \mathcal{K}_1(r) + (r^2 + R^2) \mathcal{I}_2(R) \mathcal{K}_2(r) - \frac{1}{2} r R \mathcal{I}_3(R) \mathcal{K}_3(r) \right]\left( \tilde{a}_2 \cos 2\theta + \tilde{b}_2 \sin 2\theta\right) \nonumber \\
 & +\frac{R^2}{32A} \left[ r^2 \mathcal{I}_2(R) \mathcal{K}_2(r) - 2 r R \mathcal{I}_3(R) \mathcal{K}_3(r) + R^2 \mathcal{I}_4(R) \mathcal{K}_4(r) \right] \left( \tilde{a}_2 \cos 4\theta + \tilde{b}_2 \sin 4\theta \right),
\end{align}
\begin{align}
 \tilde{u}^{(\mathrm{i})}_3 =& \frac{R^2}{384A} \left[ R^3 \mathcal{K}_1(R) \mathcal{I}_1(r) - 3 r R^2 \mathcal{K}_0(R) \mathcal{I}_0(r) + 3 R r^2 \mathcal{K}_1(R) \mathcal{I}_1(r) - r^3 \mathcal{K}_2(R) \mathcal{I}_2(r) \right] \left( \tilde{a}_2 \cos \theta - \tilde{b}_2 \sin \theta \right) \nonumber\\
 &+ \frac{R^2}{128A} \left[ -3 r R^2 \mathcal{K}_0(R) \mathcal{I}_0(r) + R(3 r^2 + R^2) \mathcal{K}_1(R) \mathcal{I}_1(r) - r^3 \mathcal{K}_2(R) \mathcal{I}_2(r) \right] \left( \tilde{a}_2 \cos \theta + \tilde{b}_2 \sin \theta\right) \nonumber\\
 &+ \frac{R^2}{128A} \left[ 2 r^2 R \mathcal{K}_1(R) \mathcal{I}_1(r) - r \left(r^2 + \frac{8}{3} R^2 \right) \mathcal{K}_2(R) \mathcal{I}_2(r) + R (r^2 + R^2) \mathcal{K}_3(R) \mathcal{I}_3(r) - \frac{1}{3}r R^2 \mathcal{K}_4(R) \mathcal{I}_4(r) \right] \nonumber \\
 &\quad \times \left( \tilde{a}_2 \cos 3\theta+ \tilde{b}_2 \sin 3\theta \right) \nonumber \\
 & + \frac{R^2}{384A} \left[ -r^3 \mathcal{K}_2(R) \mathcal{I}_2(r) + 3 r^2 R \mathcal{K}_3(R) \mathcal{I}_3(r) - 3 r R^2 \mathcal{K}_4(R) \mathcal{I}_4(r) + R^3 \mathcal{K}_5(R) \mathcal{I}_5(r) \right] \left( \tilde{a}_2 \cos 5\theta + \tilde{b}_2 \sin 5\theta \right) ,
\end{align}
\begin{align}
 \tilde{u}^{(\mathrm{o})}_3 =& \frac{R^2}{384A} \left[ R^3 \mathcal{I}_1(R) \mathcal{K}_1(r) - 3 r R^2 \mathcal{I}_0(R) \mathcal{K}_0(r) + 3 R r^2 \mathcal{I}_1(R) \mathcal{K}_1(r) - r^3 \mathcal{I}_2(R) \mathcal{K}_2(r) \right] \left( \tilde{a}_2 \cos \theta - \tilde{b}_2 \sin \theta\right) \nonumber\\
 & + \frac{R^2}{128A} \left[ -3 r R^2 \mathcal{I}_0(R) \mathcal{K}_0(r) + R(3 r^2 + R^2) \mathcal{I}_1(R) \mathcal{K}_1(r) - r^3 \mathcal{I}_2(R) \mathcal{K}_2(r) \right] \left( \tilde{a}_2 \cos \theta + \tilde{b}_2 \sin \theta\right) \nonumber\\
 & + \frac{R^2}{128 A} \left[ 2 r^2 R \mathcal{I}_1(R) \mathcal{K}_1(r) - r \left(r^2 + \frac{8}{3} R^2 \right) \mathcal{I}_2(R) \mathcal{K}_2(r) + R (r^2 + R^2) \mathcal{I}_3(R) \mathcal{K}_3(r) - \frac{1}{3} r R^2 \mathcal{I}_4(R) \mathcal{K}_4(r) \right] \nonumber \\
 & \quad \times \left( \tilde{a}_2 \cos 3\theta + \tilde{b}_2 \sin 3\theta \right) \nonumber \\
&+ \frac{R^2}{384A} \left[ -r^3 \mathcal{I}_2(R) \mathcal{K}_2(r) + 3 r^2 R \mathcal{I}_3(R) \mathcal{K}_3(r) - 3 r R^2 \mathcal{I}_4(R) \mathcal{K}_4(r) + R^3 \mathcal{I}_5(R) \mathcal{K}_5(r) \right] \left( \tilde{a}_2 \cos 5\theta + \tilde{b}_2 \sin 5\theta \right).
\end{align}
These explicit forms are obtained by the perturbation method with respect to small parameters $\epsilon$ and $v$. First, the series expansion in \eqref{us} is plugged in Eqs.~\eqref{eqS13} and \eqref{eqS14} and the equations are obtained with every order of $\epsilon$ and $v$, which are successively solved. We should consider the condition in which the concentration field $u$ and its gradient $\nabla u$ are continuous at the periphery of $\Omega( \bm{r}_c, a_k, b_k)$~\cite{Kitahata_Iida_2013,iida2014theoretical,Kitahata_JPSJ_2020,Kitahata_Frontier2022}.

\twocolumngrid

\section{Fitting of the coefficient of the cubic term of the deformation \label{app:h0}}

\begin{figure}
    \centering
    \includegraphics{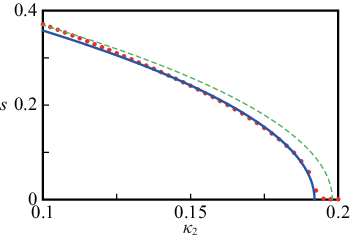}
    \caption{Magnitude $s$ of deformation depending on $\kappa_2$ obtained by the numerical simulation only considering the deformation dynamics. The simulation results are shown with red dots. The result of the fitting to a function $\kappa_2 = p_0 - p_2 s^2$ is shown with a blue line. Here, $p_0$ and $p_2$ are obtained to be $p_0 = 0.1184$ and $p_2 = 0.1436$ from the fitting. The green broken line shows the curve using the analytically estimated value for $\tilde{g}_0$. \label{figA}}
\end{figure}

Since our model started from the free energy considering the first order of the deformation magnitude, we cannot analytically derive the coefficient of the cubic term with respect to the deformation magnitude. For the ODE approach, we need to know the value of the coefficient of the cubic term. Therefore, we run the numerical simulation and obtain the coefficient by fitting the simulation data. Since we only need to consider the transition from an IC droplet to an ID droplet, we do not include the dynamics of the translational motion but only consider the deformation dynamics coupled with the time evolution of the concentration field. Other parameters are set to be the same as described in the main text. The plot of the deformation magnitude $s$ versus $\kappa_2$ is shown with red dots in Fig.~\ref{figA}.

The simulation results are fitted by the function
\begin{align}
    \kappa_2 = p_0 - p_2 s^2,
\end{align}
with the least square method. Since we need to focus on the region with small $\left|s\right|$, we only use the data with $0.001 < \left|s\right| < 0.3$. By the fitting, we obtain $p_0 = 0.1184$ and $p_2 = 0.1436$. The resulting curve is shown with a blue solid line in Fig.~\ref{figA}. From the theoretical analysis, the constant term corresponding to $p_0$ is estimated to be $\tilde{g}_0$, which is estimated as $\tilde{g}_0 \simeq 0.1196$ with the corresponding parameters. The curve using this estimated value for $\tilde{g}_0$ is also plotted by a green broken line in Fig.~\ref{figA}.
The difference between the estimated value of $\tilde{g}_0$ and $p_0$ may be due to the smoothing effect. For obtaining the phase diagram in Fig.~\ref{fig3}, we adopt the estimated value from the theory for $\tilde{g}_0$ and $p_2 \simeq 0.1436$ for $h_0$.

\section{Details on analyses of the ODE model \label{app:detail}}

First, we discuss the stability of an immobile circular (IC) droplet, $v=s=0$. Considering the linear terms of Eqs.~\eqref{eq_v} and \eqref{eq_s}, the IC droplet is stable when $\eta_t > f_1$ and $\kappa_2 > \tilde{g}_0$. In the case of $\kappa_2 < \tilde{g}_0$, the solution corresponding to an immobile deformed (ID) droplet exists: $v= 0$ and $s = \sqrt{(\tilde{g}_0 - \kappa_2)/h_0}$. This transition from IC to ID droplets is a supercritical pitchfork bifurcation.

Next, we consider a mobile deformed (MD) droplet, in which both $v$ and $s$ have finite values. From Eqs.~\eqref{eq_phi} and \eqref{eq_psi}, we obtain the equation for $\xi = \psi - \phi$ as
\begin{align}
\frac{d\xi}{dt} = \left( \frac{g_2 v^2}{2\eta_2 s} + \frac{\tilde{f}_1 s}{m}\right) \sin 2\xi.
\end{align}
Considering that the coefficient of $\sin 2\xi$ is always positive, $\xi$ converges to $\pm \pi/2$. Thus, hereafter, we assume $\xi = \pm \pi/2$ always hold. It is notable that $\xi = \pm \pi/2$ indicates that the droplet moves in its minor-axis direction. Then, the solution corresponding to an MD droplet is obtained by simultaneously solving 
\begin{align}
\left(-\eta_t + f_1 \right) v + \tilde{f}_1 s v - f_3 v^3 = 0,
\end{align}
\begin{align}
 \left(-\kappa_2 + \tilde{g}_0 \right) s + g_2 v^2  - \tilde{g}_2 s v^2 - h_0 s^3 = 0.
\end{align}
By eliminating $v$ from these two equations, we obtain the equation for $s$
\begin{align}
    H(s) =& f_3 h_0 s^3 + \tilde{f}_1\tilde{g}_2 s^2 + \left[ - f_3(\tilde{g}_0 - \kappa_2)\right. \nonumber \\
    & \left. -\tilde{g}_2 (\eta_t - f_1) - g_2 \tilde{f}_1 \right]s + g_2(\eta_t - f_1) = 0. \label{Hs}
\end{align}
The number of solutions of Eq.~\eqref{Hs} changes when $H(s)$ touches the $s$ axis. Thus, by obtaining $s$ that holds
\begin{align}
H(s) = \frac{dH}{ds} = 0, \label{funcboundary}
\end{align}
we can obtain the boundary of the region in which the solution for an MD droplet exists. 

Finally, the stability of an ID droplet is discussed. By substituting $s$ in Eq.~\eqref{eq_v} with the value of $s$ for an ID droplet, we obtain
\begin{align}
    m\frac{dv}{dt} = \left(-\eta_t + f_1 + \tilde{f}_1 \sqrt{\frac{\tilde{g}_0 - \kappa_2}{h_0}}\right)v - f_3 v^3.
\end{align}
Therefore, the threshold of the stability for an ID droplet is
\begin{align}
    \eta_t = f_1 + \tilde{f}_1 \sqrt{\frac{\tilde{g}_0 - \kappa_2}{h_0}}. \label{idmd}
\end{align}
Therefore, the curves $\eta_t = f_1$, $\kappa_2 = \tilde{g}_0$, Eq.~\eqref{funcboundary} with Eq.~\eqref{Hs}, and Eq.~\eqref{idmd} indicate the boundary of a stable IC droplet region when decreasing $\eta_t$, the boundary between stable IC and ID droplet regions, the boundary of a stable MD droplet when increasing $\eta_t$ or $\kappa_2$, and the boundary of the stable ID droplet when decreasing $\eta_t$ or $\kappa_2$, respectively, as shown in Fig.~\ref{fig3}(a).

In order to draw the bifurcation diagram, first we obtain the steady-state solution. The steady-state solutions for the IC and ID droplets are given analytically, and those for the MD droplets are given using the bisection method. For the deformed droplets, the solutions for the minor-axis and major-axis directed motion are considered; the solutions moving in the other directions should not be the steady state solution. Then, we obtain the linearized equations of Eqs.~\eqref{eq58} and \eqref{eq59}, and calculate the eigenvalues of the Jacobi matrix of the linearized equation, which gives us the information about the stability and the bifurcation type, which are shown in Fig.~\ref{fig3}(b).

\end{document}